\documentclass[letterpaper, 10 pt, conference]{ieeeconf}  % Comment this line out if you need a4paper

\IEEEoverridecommandlockouts                              % This command is only needed if 
\usepackage{graphicx} % for pdf, bitmapped graphics files
\usepackage{caption}
\usepackage{subcaption}
\usepackage{epsfig} % for postscript graphics files
\usepackage{newlfont}
\usepackage{dsfont}
\usepackage{amsmath} % assumes amsmath package installed
\usepackage{amssymb}  % assumes amsmath package installed
\usepackage{amsfonts}
\usepackage{color}

\usepackage{mathtools}

\newcommand{\qed}{\hfill $\blacksquare$}

{}
{}
\newtheorem{proposition}{Proposition}{}
\newtheorem{theorem}{Theorem}{}
{}
\newtheorem{lemma}{Lemma}{}

\DeclareMathOperator*{\argmax}{arg\,max}

\title{\LARGE \bf
Tuning Windowed Chi-Squared Detectors for Sensor Attacks
}

\author{Tunga R$^{1}$, Carlos Murguia$^{2}$, and Justin Ruths$^{1}$% <-this % stops a space
\thanks{*This work was partially supported by the National Research Foundation (NRF), Prime Minister's Office, Singapore, under its National Cybersecurity R\&D Programme (Award No. NRF2014NCR-NCR001-40) and administered by the National Cybersecurity R\&D Directorate.}% <-this % stops a space
\thanks{$^{1}$These authors are with the Departments of Electrical ({\tt\small txr160830}) and Mechanical and Systems ({\tt\small jruths}) Engineering at the University of Texas at Dallas, Richardson, Texas, USA 
        {\tt\small @utdallas.edu}}%
\thanks{$^{2}$C. Murguia is with the iTrust Centre at the Singapore University of Technology and Design, Singapore
        {\tt\small murguia\_rendon@sutd.edu.sg}}%
}

\begin{document}

\maketitle
\thispagestyle{empty}
\pagestyle{empty}

%%%%%%%%%%%%%%%%%%%%%%%%%%%%%%%%%%%%%%%%%%%%%%%%%%%%%%%%%%%%%%%%%%%%%%%%%%%%%%%%
\begin{abstract}

A model-based windowed chi-squared procedure is proposed for identifying falsified sensor measurements. We employ the widely-used static chi-squared and the dynamic cumulative sum (CUSUM) fault/attack detection procedures as benchmarks to compare the performance of the windowed chi-squared detector. In particular, we characterize the state degradation that a class of attacks can induce to the system while enforcing that the detectors do not raise alarms (zero-alarm attacks). We quantify the advantage of using dynamic detectors (windowed chi-squared and CUSUM detectors), which leverages the history of the state, over a static detector (chi-squared) which uses a single measurement at a time. Simulations using a chemical reactor are presented to illustrate the performance of our tools.
\end{abstract}

%%%%%%%%%%%%%%%%%%%%%%%%%%%%%%%%%%%%%%%%%%%%%%%%%%%%%%%%%%%%%%%%%%%%%%%%%%%%%%%%
\section{INTRODUCTION}

There has recently been significant interest and work in the broad area of security and privacy of cyber-physical systems (CPSs), see for example \cite{Cardenas}\nocite{Pasqualetti_1}\nocite{Mo_1}\nocite{Kwon}\nocite{Pappas}-\cite{Gupta2}. This topic investigates the properties of conventional control systems in the presence of adversarial disturbances. Control theory has shown great ability to robustly deal with disturbances and uncertainties. However, adversarial attacks raise all-new issues due to the aggressive and strategic nature of the disturbances that attackers might inject into the system. Moreover, large-scale control systems are often geographically distributed and so now depend heavily on communication networks for monitoring and coordinating physical processes. The use of communication networks makes these systems more prone to \textit{false data injection attacks} \cite{Liu1}, i.e., between transmission and reception of data, attackers may inject signals to data coming from the sensors/controller to the controller/process. It is therefore important to quantify what the attacker can do to the system given the process dynamics, the controller structure, and fault detection scheme. 

%Modern control systems are geographically distributed and depend on communication networks for monitoring and coordinating physical processes. The use of communication networks makes the systems more prone to attacks, i.e., between transmission and reception of data, attackers may replace the signals coming from the sensors/controller to the controller/process. It is therefore important to quantify and mitigate what the attacker can do to the system given the process dynamics and the controller structure.

In this manuscript, we focus on a class of attacks that are not detected by the given detection mechanism, which we refer to as \textit{zero-alarm attacks}. Because these attacks occur without being detected by the operator, it is important to characterize the impact that they may induce on the system performance. In particular, we consider attacks on LTI stochastic systems equipped with standard Kalman-filter based fault/attack detection procedures. The main idea behind these schemes is the use of a Kalman-filter to forecast the evolution of the system state. This prediction is compared with sensor measurements. If the difference between what it is measured and the prediction (referred to as the {residual}) is larger than expected, there might be a fault in or an attack on the system. There exist many well-known techniques which may be used to examine the residuals and subsequently detect faults/attacks. For instance, Sequential Probability Ratio Testing (SPRT) \cite{wald,Willsky}, Cumulative Sum (CUSUM) \cite{Gustafsson,Page}, Generalized Likelihood Ratio (GLR) testing \cite{Basseville}, Compound Scalar Testing (CST) \cite{Gertler}, etc. Each of these techniques has its own advantages and disadvantages depending on the scenario. Here, we focus on the a particular case of CST, the so-called \emph{windowed chi-squared} change detection procedure. The windowed chi-squared procedure considers a moving sum of a function of the residuals over a sliding window. This detector is a popular choice for fault detection and a number of groups now use it for attack detection \cite{guo2017optimal,li2017detection,wu2017optimal,mo2009secure,mo2014detecting}. It benefits from a history of observations, which makes it more adaptive than the one-shot version, the static chi-squared detector. However, as its name implies, it retains much of the analytic tractability of the static chi-squared detector, leveraging a large body of work on the chi-squared distribution. We discuss the performance of the windowed chi-squared detector in comparison to the static chi-squared and the dynamic CUSUM detectors in terms of performance degradation induced by zero-alarms attacks. The trade-offs affecting the performance of the windowed chi-squared are also explored.

\section{BACKGROUND}
We consider stochastic discrete-time linear time-invariant (LTI) systems of the form:
\begin{equation} \label{eq:dLTIsystem}
\left\{\begin{aligned}
	x_{k+1} &= Fx_k + Gu_{k} + v_{k},\\
	y_k &= Cx_k + \eta_k, \\
\end{aligned}\right.
\end{equation}
with state $x_k \in \mathbb R^{n}$, control input $u_k \in \mathbb R^{m}$, measurements $y_k \in \mathbb R^{p}$. Such systems arise, for example, from a stochastic continuous-time LTI plant sampled with time-invariant sampling ($h \in \mathbb{R}_{>0}$) at time-instants $t_{k} := kh$. The matrices $F$ and $G$, of appropriate dimensions, are then the sampling dependent matrices representing the discrete-time state and input matrices, respectively. The $C$ matrix captures the mapping from states to measurements. The vector $v_k \in \mathbb R^{n}$ represents additive system noise, i.i.d zero-mean Gaussian random noise with covariance matrix $R_1$ and $\eta_{k} \in \mathbb R^{p}$ represents measurement noise, i.i.d. zero-mean Gaussian random noise with covariance matrix $R_2$. We assume that the pair $(F,C)$ is detectable and $(F,G)$ is stabilizable. 

In this work, we consider the scenario that the actual measurement $y_k$ can be corrupted by an additive attack, $\delta_k \in \mathbb{R}^p$. At some point in the process of measuring and transmitting the output to the controller the attacked output becomes
\begin{equation}
	\bar{y}_k = y_k + \delta_{k} =  Cx_{k} + \eta_{k} + \delta_k.
\end{equation}
If the attacker has access to the measurements, then it is possible for the attack $\delta_k$ to cancel some or all of the original measurement ${y}_k$ - so an additive attack can achieve arbitrary control over the ``effective'' output of the system.

As our approach leverages a fault-detection approach, we require an estimator of some type to produce a prediction of the system behavior. In this work we use the steady state Kalman filter
\begin{equation}
	\hat{x}_{k+1} = F\hat{x}_k + Gu_k + L(\bar{y}_k - C\hat{x}_k),
\end{equation}
where $\hat{x}_k \in \mathbb{R}^n$ is the estimated state. The observer gain $L$ is designed to minimize the steady state covariance matrix $P:= \lim_{k \rightarrow \infty}P_k:= E[e_ke_k^T]$ in the absence of attacks, where $e_k:= x_k - \hat{x}_k$ denotes the estimation error. Existence of $P$ is guaranteed since the pair $(F,C)$ is assumed to be detectable \cite{Astrom}. Next, we define the residual sequence $r_k$
\begin{equation}
	r_k := \bar{y}_k - C\hat{x}_k,
\end{equation}
which evolves according to
\begin{equation} \label{eq:esterror}
\left\{ \begin{aligned}
	e_{k+1} &= \big( F - LC \big) e_k - L\eta_k + v_k - L \delta_k, \\
	r_k &= Ce_k + \eta_k + \delta_k.
\end{aligned}\right.
\end{equation}
In the absence of attacks (i.e., $\delta_k = 0$), it is straightforward to show that the $r_k$ random variable falls according to a zero mean Gaussian distribution with covariance \cite{Carlos_Justin2}
\begin{equation}
	\Sigma = E[r_kr_k^T] = CPC^T+R_2.
\end{equation}
Anomaly detection relies on identifying deviations from this nominal distribution. While in theory it is possible to aggregate enough samples to reconstruct a sample distribution and compare this with the nominal distribution, this approach would be extremely slow. Instead, detectors are designed to make decisions to raise alarms based on much less information.

\subsection{Detection procedures}

An anomaly (including faults and attacks) in a dynamical system can be defined as a deviation of measurements away from those predicted by the estimated state. Anomaly detection uses the residual $r_k$ (difference between what is measured and estimated). If the residual is larger than expected, or specifically larger than a decision threshold, there might be a fault in or attack on the system. Several types of detectors exist; this work extends our past findings to the popular windowed chi-squared detector. 

The performance of a detector is related to its ability to correctly identify true anomalies (quantified by the true positive rate) and its ability to avoid raising alarms when there are no anomalies (quantified by the false alarm rate). Understanding the trade-off of these two rates is critical to assessing the quality of a detector. Toward this aim, it is important to be able to tune detectors to achieve different points along the receiver operating characteristic curve. This is one of the main outcomes of our past work and also what we contribute in this paper with respect to the windowed chi-squared detector. To help quantify these values, we define the \textit{run length} $K$ as the number of measurements needed (without anomalies present) such that the detector raises an alarm:
\begin{equation}
	K := \min\{{k \geq 1 : S_k > \tau}\},
\end{equation}
where $S_k$ denotes the decision variable of the detector. The expected value $E[K]$ is known as the \textit{Average Run Length} (ARL) and is inversely proportional to the false alarm rate of the detector $\mathcal{A}$, i.e., $\mathcal A = 1/ARL$.

\subsubsection{Static chi-squared detector} 
The chi-squared detector uses a quadratic form of the residual to test for substantial variations in the covariance and expected value of the error between the observed and estimated outputs. This quadratic so-called \textit{distance measure} has several advantages over the arguably simplest detector that simply compares the absolute error to a threshold. Beyond testing for changes in the spread of the residual distribution, the chi-squared distance measure $z_k = r_k^T\Sigma^{-1}r_k$ provides an analytically tractable distribution. Since $r_k \sim \mathcal N(0,\Sigma)$, the $z_k$ random variable, as the sum of the squares of normally distributed random variables, falls according to the chi-square distribution. Since $r_k\in\mathbb{R}^p$, this chi-squared distribution has $p$ degrees of freedom. The chi-squared detector is summarized as follows: for given a threshold $\alpha \in \mathbb{R}_{>0}$ and the distance measure \linebreak $z_k = r_k^T\Sigma^{-1}r_k$
\begin{equation} 
\left\{\begin{aligned}
	z_k \leq \alpha &\quad\xrightarrow{\text{no alarm}}\quad \\
	z_k > \alpha &\quad\xrightarrow{\text{alarm}\ \ \; }\quad k^* = k,
\end{aligned}\right.
\end{equation}
alarm time(s) $k^*$ are produced. The $\Sigma^{-1}$ factor in the definition of $z_k$ rescales the distribution ($E[z_k]=p$, $E[z_kz_k^T]=2p$) so that the threshold $\alpha$ can be designed independent of the specific statistics of the noises $v_k$ and $\eta_k$; instead it can be selected simply based on the number of sensors (i.e., the dimension of the output, $p$). The following lemma indicates how to select the threshold $\alpha$ to achieve a desired rate of false alarms.

\begin{lemma} \cite{Carlos_Justin2}. \label{lem:chisquared_tuning}
Assume that there are no attacks to the system and consider chi-squared detector, with threshold $\alpha \in \mathbb{R}_{>0}$, $r_k \sim N(0,\Sigma)$. Let $\alpha = \alpha^{*}:=2P^{-1}(1-\mathcal A^{*}, \frac{p}{2})$, where $P^{-1}(\cdot,\cdot)$ denotes the inverse regularized lower incomplete gamma function, then ${\mathcal A}= {\mathcal A}^{*}$. 
\end{lemma} 

%The quadratic expression $r_k^T\Sigma^{-1}r_k$ represents the distance measure where $r_k$ and $\Sigma$ are the residual sequence and covariance matrix respectively. 
%\begin{equation}
%\left\{
%\begin{array}{lc}
%E[z_k] = m\\
%var[z_k] = 2m\\
%\end{array}
%\right.
%\end{equation}

\subsubsection{CUSUM detector}
CUSUM (Cumulative Sum) is a sequential analysis technique which is used to aggregate the error over an adaptive window to protect the system against small, but persistent, attacks \cite{Page}. Given a chosen distance measure $z_k$ (here we consider the quadratic distance measure $z_k = r_k^T\Sigma^{-1}r_k$, but it is possible to drive the CUSUM with any other choice of distance measure), the CUSUM procedure evolves as follows: for given threshold $\tau\in\mathbb{R}_{>0}$, bias parameter $b \in\mathbb{R}_{>0}$, and initial cumulative sum $S_1 = 0$ 
\begin{equation} \label{eq:CUSUM}
\left\{\begin{aligned}
    &S_{k-1}\le\tau \quad\xrightarrow{\text{no alarm}}\quad S_k = \max(0, S_{k-1}+z_k-b) \\
	&S_{k-1}>\tau \quad\xrightarrow{\text{\hspace{2mm}alarm} \ }\quad S_k = 0,\ k^* = k-1.
\end{aligned}\right.
\end{equation}
Effectively the test sequence $S_k$ accumulates the distance measure $z_k$ and alarms the triggered when $S_k$ exceeds the threshold $\tau$. The test is reset to zero each time $S_k$ becomes negative or exceeds $\tau$ \cite{Page}. Since $z_k \geq 0$, the bias parameter $b$ prevents the inherent growth due to the sum of a nonnegative number. Intuitively, the $z_k$ distribution should be shifted by its mean, i.e., $b \geq E[z_k] = p$ such that $E[z_k-b]\leq 0$; this is proved rigorously in \cite{Carlos_Justin2}. For the tightest detection $b\approx p$. This same article provides similar selection of the threshold $\tau$ to achieve a desired false alarm rate as present above for the chi-squared detector, however, due to the nonlinear dynamics of the CUSUM statistic \eqref{eq:CUSUM} there is no closed form solution (however, the proposed heuristics can be made arbitrarily accurate).

%  \begin{lemma}[\cite{Carlos_Justin2}]
% Assuming that there are no attacks to the system, i.e. $\delta_{\textit{k}} = 0$, let the CUSUM with bias $b \in \mathbb{R}_{>0}$ and threshold $\tau \in \mathbb{R}_{>0}$ be driven by the distance measure ${z_k} = {r_k}^T\Sigma^{-1}{r_k}$, $\textit{k} \in \mathbb{N}$ with residual sequence $r_k\sim \mathcal N(0,\Sigma)$, $\textit{k} \in \mathbb{N}$. Then for $b > \bar{b}:=p$, the CUSUM sequence $S_k, k \in \mathbb{N}$ is bounded in mean square sense independent of the threshold. 
% \end{lemma} 

\subsection{Zero alarm attacks}
Zero alarm attacks, introduced in \cite{Carlos_Justin2},\cite{Carlos_Justin1}, represent a powerful attacker model which allows us to characterize the worst case steady state impact an attacker can have. Zero alarm attacks are then useful for system design because it allows operators to understand the extent of damage an attacker can cause if the attacker chooses to remain beneath the threshold of detection.  While other attack models are possible zero alarm attacks provide a useful and clear benchmark.  Hidden attacks, which allow attackers to raise alarms as long as the alarm rate matches the false alarm rate of the detector, are also a useful benchmark, however, we have shown that such attacks can lead to arbitrarily large state deviations unless additional detection methods are implemented \cite{Carlos_Justin3,Mo_3}. Additional detection methods then confuse the original objective of our work: to assess how various detectors yield different attacker capabilities. Zero alarm attacks are designed to maintain the entire attacked test statistic (e.g., $z_k$ for a chi-squared detector; $S_k$ for a CUSUM detector) distribution at or below the detector threshold ($\alpha$ and $\tau$, respectively), effectively maximizing the false negative rate (since the true positive rate is zero, i.e., the true attack is never detected).

In order to observe the effect of the attack on the system state (not just on the estimation error), we need to close the control loop with a controller. Here we consider the output feedback controller $u_k = K\hat{x}_k$, where $\hat{x}_k \in \mathbb{R}^{n}$ is the state of Kalman filter and $K \in \mathbb{R}^{m \times n}$ denotes the feedback control matrix. Then, the closed-loop system becomes
\begin{equation}
\left\{\begin{aligned} \label{eq:state_error}
	x_{k+1} &= (F+GK)x_k+G K e_k+v_k,\\
	e_{k+1} &= (F-LC)e_k-L\delta_k-L\eta_k+v_k.\\
\end{aligned}\right.
\end{equation}

When zero alarm attacks are designed against the chi-squared detector the attack sequence becomes
\begin{equation}\label{eq:zero_alarm_chisquared}
	\delta_k = C\hat{x}_k-y_k + \Sigma^{\frac{1}{2}}\bar{\alpha} = - Ce_k-\eta_k+\Sigma^{\frac{1}{2}}\bar{\alpha},
\end{equation}
where $\bar\alpha\in\mathbb{R}^p$ is any vector such that $\bar\alpha^T\bar\alpha = \alpha$. Notice that this attack either assumes the attacker has access to $y_k$ or has the ability to replace the original measurement with another value. The technology to inject a sensor attack typically also enables the attacker to ``sniff'' communication packets (know measurement values) and manipulate packets to add to the value they carry or completely replace it. Thus these are reasonable assumptions for quantifying worst case attacks. This attack sequence leads the distance measure to become
\begin{align*}
	z_k &= r_k^T\Sigma^{-1}r_k \\
    &= (Ce_k+\eta_k+\delta_k)^T\Sigma^{-1}(Ce_k+\eta_k+\delta_k) \\
    &= \alpha,
\end{align*}
which does not raise alarms by the chi-squared detector with threshold $\alpha$.
% \begin{equation}
% \left\{
% \begin{array}{lcc}
% E[x_{k+1}] = (F+GK)x_k-GKE[e_k],\\
% E[e_{k+1}] = FE[e_k]-L\Sigma^{1/2}\bar{\alpha}\\
% \end{array}
% \right.
% \end{equation}

\begin{lemma} \cite{Carlos_Justin2}.
Consider the chi-square detector and let the sensors be attacked by the chi-squared zero-alarm attack sequence \eqref{eq:zero_alarm_chisquared}. If $\rho{[F]}<1$, where $\rho[\cdot]$ denotes spectral radius, then $\lim_{k \to \infty} \|E[x_k]\|=\gamma_{\chi^{2}}$, where
\begin{equation} \label{eq:gamma_chisquared}
\gamma_{\chi^{2}}:= \Big\|(I-F-GK)^{-1}GK(I-F)^{-1}L\Sigma^{\frac{1}{2}}\bar{\alpha}\Big\|.
\end{equation}
\end{lemma}
\vspace{2ex}

Similarly, for the CUSUM detector, the zero-alarm attack sequence is given by
\begin{equation} \label{eq:zero_alarm_cusum}
	\delta_k = \left\{
    \begin{aligned}
    	-Ce_k-\eta_{k}+\Sigma^{1/2}\bar{\tau},\qquad & k=k^{*}, \\
        -Ce_k-\eta_{k}+\Sigma^{1/2}\bar{b},\qquad & k>k^{*}, 
    \end{aligned}\right.
\end{equation}
where $\bar{b} \in\mathbb{R}^p$ (resp., $\bar{\tau} \in\mathbb{R}^p$) is any vector such that $\bar{b}^T\bar{b} = b$ (resp., $\bar{\tau}^T\bar{\tau} = \tau$) and $k^*$ is the starting attack instant. The first step of this attack sequence saturates the test statistic $S_k=\tau$ and the subsequent steps maintain the statistic at the threshold.

\begin{lemma} \cite{Carlos_Justin2}.
Consider the CUSUM detector and let the sensors be attacked by the CUSUM zero alarm attack sequence \eqref{eq:zero_alarm_cusum}. If $\rho{[F]}<1$, then $\lim_{k \to \infty} \|E[x_k]\|=\gamma_{\text{CS}}$, where
\begin{equation} \label{eq:gamma_cusum}
\gamma_\text{CS}:= \Big\|(I-F-GK)^{-1}GK(I-F)^{-1}L\Sigma^{\frac{1}{2}}\bar{b}\Big\|.
\end{equation}
\end{lemma}
\vspace{1ex}

\section{WINDOWED CHI-SQUARED DETECTOR}
Our past work has identified how to tune the chi-squared and CUSUM detectors to achieve desired rates of false alarm. We also quantified the effect that attackers can have on the system state while remaining underneath the detection limit of these detectors. While chi-squared represents one of the most popular detector choices in literature due to its simplicity and analytic tractability, our past work has also shown some advantages of the dynamic CUSUM detector to maintain sensitivity to low-amplitude but long-term attacks. Because the CUSUM bias $b$ is typically selected smaller than the chi-squared threshold $\alpha$, the attacker capabilities are much reduced using a CUSUM detector since $b<\alpha$ implies $\gamma_\text{CS} < \gamma_{\chi^2}$.

A compelling middle-ground option is provided by the windowed chi-squared detector, which incorporates a sliding window that sums the distance measure over the window interval. This detector maintains most, if not all, of the analytic tractability of the (static) chi-squared detector, while providing enhanced detector sensitivity. On a theoretic level it provides a way to understand the effect of the length of the window on attacker capabilities, since a window of length one recovers the static chi-squared and letting the window length grow longer resembles some aspects of the CUSUM detector (without the reset characteristic). Formally, the windowed chi-squared detector is summarized as follows: for given threshold $\beta\in\mathbb{R}_{>0}$, distance measure $z_k = r_k^T\Sigma^{-1}r_k$, and test statistic
\begin{equation} 
	w_k = \sum_{i = k-l+1}^{k} z_i = \sum_{i = k-l+1}^{k} r_i^T\Sigma^{-1}r_i,
\end{equation}
the detector compares
\begin{equation} \label{eq:windowedchisquared}
\left\{\begin{aligned}
    w_k \leq \beta &\quad\xrightarrow{\text{no alarm}}\quad \\
	w_k > \beta &\quad\xrightarrow{\text{alarm}\ \ \; }\quad k^* = k.
\end{aligned}\right.
\end{equation}
Here, the moving sum of $z_i$ is taken over a sliding window $[k-\ell+1,k]$ of length $\ell$. The extension to appropriately tune the windowed chi-squared detector is similar to Lemma \ref{lem:chisquared_tuning}.\\

\begin{theorem}
Assume that there are no attacks to the system and consider the windowed chi-squared detector with threshold $\beta \in \mathbb{R}_{>0}$ and $r_k \sim \mathcal N(0,\Sigma)$. Let $\beta = \beta^{*}:=2P^{-1}(1-\mathcal A^{*},\frac{p\ell}{2})$, where $P^{-1}(\cdot,\cdot)$ denotes the inverse regularized lower incomplete gamma function, then $\mathcal{A}= \mathcal{A}^{*}$. 
\end{theorem}

\vspace{1ex}
\textit{Proof}: %$\widetilde{K}$ denotes the run length of the windowed chi-squared procedure, which is defined as the number of iterations needed such that $k=\widetilde{K}$ implies $w_k>\beta$ when there are no attacks. Recall the Average Run Length is given by $\text{ARL}=E[\widetilde{K}]$ and then the false alarm rate is $\mathcal A = 1/\text{ARL}$. The random variable $\widetilde{K}$ follows a geometric distribution, therefore, $\text{ARL} = E[\widetilde{K}]= 1/pr(w_k>\beta)$ and $\mathcal A=pr(w_k>\beta)$. 
By construction, for all $i\in \mathbb{N}$, $z_i$ are independent chi-squared variables with $p$ degrees of freedom. Thus, $w_k$ is the sum of $\ell$ independent chi-squared variables. The sum of two independent chi-squared variables with \textit{p} and \textit{q} degrees of freedom, respectively, is a chi-squared variable with \textit{p+q} degrees of freedom \cite {lancaster1969chi}. Hence, the windowed chi-squared statistic $w_k$ falls according to a chi-squared distribution with $p\ell$ degrees of freedom. The cumulative distribution function is then given by $\mathcal{F}(w) = P(\frac{p\ell}{2},\frac{w}{2})$, where $P(\cdot,\cdot)$ is the regularized lower incomplete gamma function. The false alarm rate is simply the portion of the $w_k$ distribution that falls in the tail beyond the threshold $\beta$
\begin{equation}
	\mathcal A = \text{pr}(w_k>\beta)= 1-\mathcal{F}(\beta) = P\left(\frac{p\ell}{2},\frac{\beta}{2}\right).
\end{equation}
Inverting this relationship to solve for $\beta$ yields the result. \qed 

Next, to quantify an attacker's capabilities under a windowed chi-squared detector, we formulate the zero-alarm attack sequence. Recall that a zero-alarm attack devises an attack sequence such that the test statistic saturates and is maintained at the detection threshold, i.e., for the windowed chi-squared detector we design $\delta_k$ such that $w_k=\beta$. Recall the definition of the windowed chi-squared statistic
\begin{equation}
	w_k = \sum_{i = k-\ell+1}^{k}{(Ce_i+\eta_i+\delta_i)^T\Sigma^{-1}(Ce_i+\eta_i+\delta_i).}
\end{equation}
To saturate $w_k = \beta$, we select $\delta_i = -Ce_i-\eta_i+\Sigma^{\frac{1}{2}}\bar{\delta}_i$, then
\begin{equation} \label{eq:zero_alarm_windowed1}
	w_k = \sum_{i = k-\ell+1}^{k} \bar\delta_i^T\bar\delta_i = \beta.
\end{equation}
This characterization of $\bar\delta_k$ leaves some ambiguity as did our definitions of $\bar{\alpha}$, $\bar{b}$ and $\bar{\tau}$ in \eqref{eq:zero_alarm_chisquared} and \eqref{eq:zero_alarm_cusum}. Because the windowed chi-squared detector is summed over the $\ell$-length time window, it is possible to design an infinite array of different $\bar\delta_k$ profiles such that $w_k=\beta$. For example the following time-varying attack sequence satisfies $w_k=\beta$:
\begin{equation}
	\bar\delta_k = 
    \begin{cases} 
    	\bar\beta, & (k-k^*)\bmod \ell = 0,\\
        0, &\text{otherwise},
    \end{cases}
\end{equation}
where
\begin{equation}\label{eq:betabar}
	\bar{\beta}:=\left\{{\bar{\beta}\in \mathbb{R}^{p}|\bar{\beta}^T\bar{\beta}=\beta}\right\}.
\end{equation}
However, both chi-squared and CUSUM detectors implement an attack sequence that does not vary with time. Therefore, to maintain an equitable comparison between the detectors, we select a static attack sequence
\begin{equation}\label{eq:static_delta}
	\bar\delta_k = \bar\delta = \frac{\bar\beta}{\ell},
\end{equation}
with $\bar\beta$ as defined in \eqref{eq:betabar}. Empirically, we observe that this static attack generates the largest steady-state state deviation, however, we leave this proof for future work. With this attack, the closed-loop dynamics \eqref{eq:state_error} becomes
\begin{equation}
\left\{\begin{aligned} \label{eq:state_error_attacked}
	x_{k+1} &= (F+GK)x_k+G K e_k+v_k,\\
	e_{k+1} &= F e_k - L\Sigma^{\frac{1}{2}}\frac{\bar\beta}{\ell} - L\eta_k + v_k.\\
\end{aligned}\right.
\end{equation}

\begin{theorem}
Consider the windowed chi-squared detector and let the sensors be attacked by the zero-alarm attack sequence \eqref{eq:static_delta}. If $\rho{[F]}<1$, then $\lim_{k \to \infty} \|E[x_k]\|=\gamma^{\ell}_{\chi^{2}}$, where
\begin{equation} \label{eq:gamma_windowed_chisquared}
\gamma^{\ell}_{\chi^{2}}:= \bigg\|(I-F-GK)^{-1}GK(I-F)^{-1}L\Sigma^{\frac{1}{2}}\frac{\bar{\beta}}{\ell}\bigg\|,
\end{equation}
and $\bar\beta$ as defined in \eqref{eq:betabar}.
\end{theorem}

\vspace{1ex}
\textit{Proof:} 
By construction and assumption, $\rho[F+GK]<1$ and $\rho[F]<1$. This implies that $(I-F-GK)$ and $(I-F)$ are invertible, respectively; hence, system (\ref{eq:state_error_attacked}) has a unique equilibrium, in expectation, given by
\begin{equation*}
\left\{
\begin{aligned}
E[{x}_{k}] &= \bar{x}:= (I - F - GK)^{-1} GK (I-F)^{-1}L \Sigma^{\frac{1}{2}} \bar{\delta},\\
E[e_{k}] &= \bar{e}:= (F-I)^{-1}L \Sigma^{\frac{1}{2}} \bar{\delta}.
\end{aligned}
\right.
\end{equation*}
Substituting \eqref{eq:static_delta} yields the matrix in $\gamma^\ell_{
\chi^2}$. To ensure this equilibrium is attractive, we use (\ref{eq:state_error_attacked}) to show that the evolution of the differences $E[e_{k}] - \bar{e}$ and $E[{x}_{k}] - \bar{x}$ satisfy
\begin{align*}
&E[{x}_{k+1}] - \bar{x}= (F + GK) (E[x_k] - \bar{x}) - GK ( E[e_k] - \bar{e} ), \\[.5mm]
&E[e_{k+1}] - \bar{e} = F (E[e_k] - \bar{e}).
\end{align*}
Since $\rho[F+GK]<1$ and $\rho[F]<1$, the equilibrium $[\bar{x},\bar{e}]^T$ is exponentially stable, i.e., $\lim_{k \rightarrow \infty} E[e_{k}] = \bar{e}$ and $\lim_{k \rightarrow \infty} E[x_{k}] = \bar{x}$. The Euclidean norm on $\mathbb{R}^n$ is a continuous\- function from $\mathbb{R}^n$ to $\mathbb{R}_{\geq 0}$ \cite{Horn}. It follows that $\lim_{k \rightarrow \infty} \|E[x_{k}]\| = \|\lim_{k \rightarrow \infty} E[x_{k}]\| =\|\bar{x}\|$. \qed

\section{WORST CASE STEADY-STATE DEVIATION}
In previous work, we identified the steady-state state deviation due to aggressive zero-alarm attacks for the chi-squared and CUSUM detectors. Here we developed an analogous result for the windowed chi-squared detector. The static nature of the zero-alarm attacks creates a similar structure for all of these steady-state bounds.  In particular if we define the matrix 
\begin{equation}\label{eq:M}
	{M}:=  (I-F-GK)^{-1}GK(I-F)^{-1}L\Sigma^{\frac{1}{2}}  \in \mathbb{R}^{n \times p},
\end{equation}
we observe the bounds can be expressed as $\gamma = \|M\bar\delta\|$, where $\bar\delta$ becomes $\bar\delta=\bar\alpha$ for the (static) chi-squared detector, $\bar\delta=\bar{b}$ for the CUSUM detector, and $\bar\delta=\bar\beta/\ell$ for the windowed chi-squared detector with window length $\ell$. The definitions of $\bar\alpha$, $\bar{b}$, and $\bar\beta$ only stipulate that their respective norms should be $\alpha$, $b$, and $\beta$. In the following result we identify, based on the matrix $M$, the ``direction'' $\bar\delta^*\in\mathbb{R}^p$ that yields the worst (largest) steady-state state deviation. In this context of attacks, this provides the relative weighting (susceptibility) of each sensor to achieve the attack with the most damage.

\begin{proposition} \label{prop:worst_attack}
Let $\lambda_1$ denote the largest eigenvalue of the positive semidefinite matrix ${M}^T {M}$ with corresponding unit eigenvector $\nu_1$. Then:\\ 
$\small{\bullet} \hspace{.5mm} \bar{\alpha} = \bar{\alpha}^* := \sqrt{\alpha} \nu_1$ satisfies $\bar{\alpha}^T \bar{\alpha} = \alpha$ and maximizes $\gamma_{\chi^2}$.
$\small{\bullet} \hspace{.5mm} \bar{b} = \bar{b}^* := \sqrt{b} \nu_1$ satisfies $\bar{b}^T \bar{b} = b$ and maximizes $\gamma_{\text{CS}}$.
$\small{\bullet} \hspace{.5mm} \bar{\beta} = \bar{\beta}^* := \sqrt{\beta} \nu_1$ satisfies $\bar{\beta}^T \bar{\beta} = \beta$ and maximizes $\gamma_{\chi^2}^\ell$.

\end{proposition}
\vspace{1ex}
\textit{Proof}: We prove this result in the context of the static chi-squared, however, the proof is identical for the other detectors. The asymptotic bound $\gamma_{\chi^2}$ can be written in terms of ${M}$ as $\gamma_{\chi^2} = \|{M}\bar{\alpha}\|$. Then, to maximize $\gamma_{\chi^2}$, we have to find the $\bar{\alpha}$ solution of the optimization problem:
\begin{equation}\label{equiv_opt2}
\begin{array}{ll}
\argmax\limits_{\bar{\alpha}^T\bar{\alpha} = \alpha} \|{M}\bar{\alpha}\| = \argmax\limits_{\bar{\alpha}^T\bar{\alpha} = \alpha} \big(\bar{\alpha}^T{M}^T{M}\bar{\alpha}\big)^{1/2}, 
\end{array}
\end{equation}
which has the same solution $\bar{\alpha} \in \mathbb{R}^m$ as
\begin{equation}\label{equiv_opt}
\argmax\limits_{\bar{\alpha}^T\bar{\alpha} = \alpha} \bar{\alpha}^T M^T M\bar{\alpha}.
\end{equation}
The matrix ${M}^T{M}$ is positive semidefinite by construction; hence, it is symmetric and its eigenvalues, $\lambda_i$, are nonnegative and real. Let the eigenvalues be ordered as $\lambda_1 \geq \lambda_2 \geq \cdots \geq \lambda_p$. This means we can write $\bar\alpha$ in terms of the orthonormal eigenvectors $\nu_i$ of $M^TM$:
\begin{equation}
	\bar\alpha = \sum_{i=1}^p a_i \nu_i\quad \rightarrow\quad
    M^TM\bar\alpha = \sum_{i=0}^p a_i \lambda_i \nu_i,
\end{equation}
where $a_i$ are the coefficients of the expansion with $\bar\alpha^T\bar\alpha  = \alpha = \sum a_i^2$. Exploiting the orthonormality of the eigenvectors, we see that
\begin{equation}
	\bar\alpha^T M^TM\bar\alpha = \sum_{i=0}^p a_i \lambda_i \bar\alpha^T\nu_i = \sum_{i=0}^p a_i^2 \lambda_i.
\end{equation}
Since $\sum a_i^2 = \alpha$ and $\lambda_1$ is the largest eigenvalue, the choice of $a_i$ coefficients that maximize this objective sets $a_1=\sqrt{\alpha}$ and $a_i=0$, $i=2,\dots,p$. This makes $\bar\alpha^* = \sqrt{\alpha}\nu_1$. \qed

% The matrix ${M}^T{M}$ is positive semidefinite by construction; hence, it is symmetric and its eigenvalues are nonnegative and real. Then, from the spectral theorem \cite{Horn}, there exists an orthonormal matrix of eigenvectors $Q \in \mathbb{R}^{p \times p}$ such that $Q^T {M}^T{M} Q = \Lambda$ where $\Lambda := \text{diag}(\lambda_1,\ldots,\lambda_p)$ and $\lambda_i \in \mathbb{R}_{\geq 0}$ denote the eigenvalues of ${M}^T{M}$, $i=1,\ldots,p$. Define the change of coordinates $\zeta := (1/\sqrt{\alpha})Q^T\bar{\alpha}$. Then, using $Q^{-1}=Q^T$ and $Q^TQ = I_p$ (because $Q$ is orthonormal), it can be verified that
% \begin{align*}
% \begin{array}{ll}
% \max\limits_{\bar{\alpha}^T\bar{\alpha} = \alpha} \bar{\alpha}^T{M}^T{M}\bar{\alpha} &= \max\limits_{\zeta^T\zeta = 1} \alpha \zeta^T \Lambda \zeta\\
% &= \max\limits_{\sum_{i=1}^p \zeta_i^2=1} \alpha \sum_{i=1}^p \lambda_i \zeta_i^2\\
% &= \alpha \max\limits_{j}  \lambda_j = \alpha \lambda_{1}[{M}^T{M}],
% \end{array}
% \end{align*}
% with corresponding index $j^*$. The corresponding optimal $\zeta^*$ has entries $\zeta_{j^*} = 1$ and $\zeta_{j \neq j^*} = 0$. Using $\zeta = (1/\sqrt{\alpha})Q^T\bar{\alpha}$, we conclude $\bar{\alpha}^* = \sqrt{\alpha}Q\zeta^* = \sqrt{\alpha} \nu^*$. \qed

\subsection{Detector comparison}
The formulations of the steady-state expectation of the state deviation in \eqref{eq:gamma_chisquared}, \eqref{eq:gamma_cusum}, and \eqref{eq:gamma_windowed_chisquared} along with the common structure exploited in Proposition \ref{prop:worst_attack} now permit us to easily make comparisons between these three detectors. In particular, we define $\bar\alpha$, $\bar{b}$, and $\bar\beta$ as in Proposition \ref{prop:worst_attack}, which identifies that $\gamma_{\chi^2}$, $\gamma_{\text{CS}}$, and $\gamma_{\chi^2}^\ell$ are different only due to the differing values of $\sqrt{\alpha}$, $\sqrt{b}$, and $\sqrt{\beta}/\ell$.

Typically low false alarm rates $\mathcal{A}$ are selected to limit the number of alarms raised in the attack-free case.  In this context, $\alpha>b\approx p$. At the same time, $\beta/\ell < \alpha$ (if $\beta/\ell \geq \alpha$ there would be no advantage of using a window length greater than 1). With these assumptions, we observe an ordering of state deviation: $\gamma_{\chi^2}>\gamma_{\text{CS}}$ and $\gamma_{\chi^2} > \gamma_{\chi^2}^\ell$. The windowed chi-squared detector performs an interesting function role and to better determine its performance against the CUSUM, we develop the following result.

% \begin{figure}[!t]
%  \begin{center}		
%  \includegraphics[width=\linewidth]{fixed_false_alarm_edited.pdf}
%  \caption{Evolution of threshold $\alpha$ with increasing window lengths, for fixed false alarm rates}
%  \end{center}           
%  \end{figure}

\begin{figure}[t]
\centering
\includegraphics[width=\linewidth]{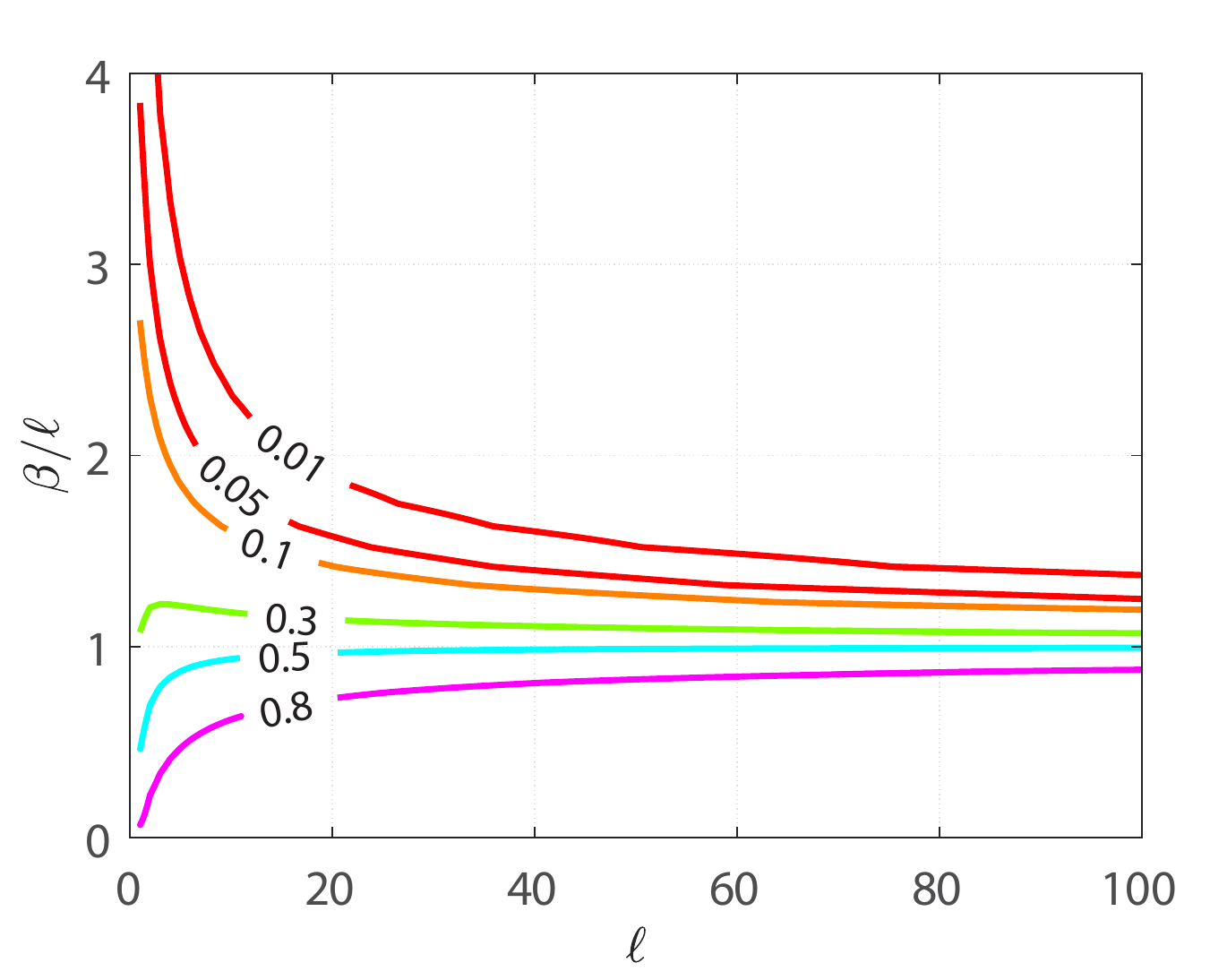}
\caption{The contour curves indicate values of window length $\ell$ and state deviation $\beta/\ell$ that correspond to equivalent false alarm rates $\mathcal{A}\in\{1\%, 5\%, 10\%, 30\%, 50\%, 80\% \}$. As the window length increases the damage (state deviation), for all false alarm rates, converges to $p$ (here $p=1$).} \label{fig:damage_vs_ell}
\end{figure}

\begin{proposition}
Given a windowed chi-squared detector with window length $\ell$ and a CUSUM detector with bias tuned to $b = p$, the following is satisfied:
\begin{equation}
	\lim_{\ell\to\infty} \gamma_{\chi^2}^\ell = \gamma_{\text{CS}},
\end{equation}
where $p$ is the dimension of the measurement vector.
\end{proposition}
\vspace{1ex}
\textit{Proof}: 
The central limit theorem provides the asymptotic properties of the sample average $S_\ell = \frac{1}{\ell}(X_1+X_2+\cdots+X_\ell)$, given a sequence of $\ell$ i.i.d. random variables, $X_1,\ X_2,\ \dots,\ X_\ell$, each with expected value $\mu$ and finite variance $\sigma^2$. The central limit theorem states that $\sqrt{\ell}(S_\ell-\mu)$ converges in distribution to $N(0,\sigma^2)$ as $\ell$ approaches infinity. 

Here, the windowed chi-squared procedure computes then the sum (not the average) of chi-squared random variables $X_k=z_k=r_k^T\Sigma^{-1}r_k$ with $p$ degrees of freedom, since $r_k\sim N(0,\Sigma)$ and $r_k\in\mathbb{R}^p$. Since $X_k=z_k$ is chi-square distributed with $p$ degrees of freedom, it has mean $\mu=p$ and variance $\sigma^2 = 2p$. Thus, by the central limit theorem, the sample average approaches $N(\mu,\frac{\sigma^2}{\ell})=N(p,\frac{2p}{\ell})$. From this, the sum (not average) approaches $N(\ell\mu,\ell\sigma^2)=N(p\ell,2p\ell)$. 

Note that determining the threshold $\beta$ to satisfy a false alarm rate $\mathcal{A}$ for the asymptotic sum distribution $N(p\ell,2p\ell)$ is equivalent to identifying the threshold $\beta/\ell$ to satisfy the same false alarm rate for the asymptotic sample average distribution $N(p,\frac{2p}{\ell})$. Notice that the latter distribution approaches mean $p$ with shrinking variance. In the limit, the variance goes to zero, which means the threshold (in fact the entire distribution) collapses down to the mean value $p$. Thus for large values of window length $\ell$, the value of $\beta/\ell$ to satisfy a chosen false alarm rate (in fact \textit{any} alarm rate) converges to $p$. 

Substituting $\beta/\ell\to p$ into $\gamma_{\chi^2}^\ell$ and substituting $b\approx p$ into $\gamma_{\text{CS}}$ then yields the result. \qed

\begin{figure}[t]
\centering
\includegraphics[width=\linewidth]{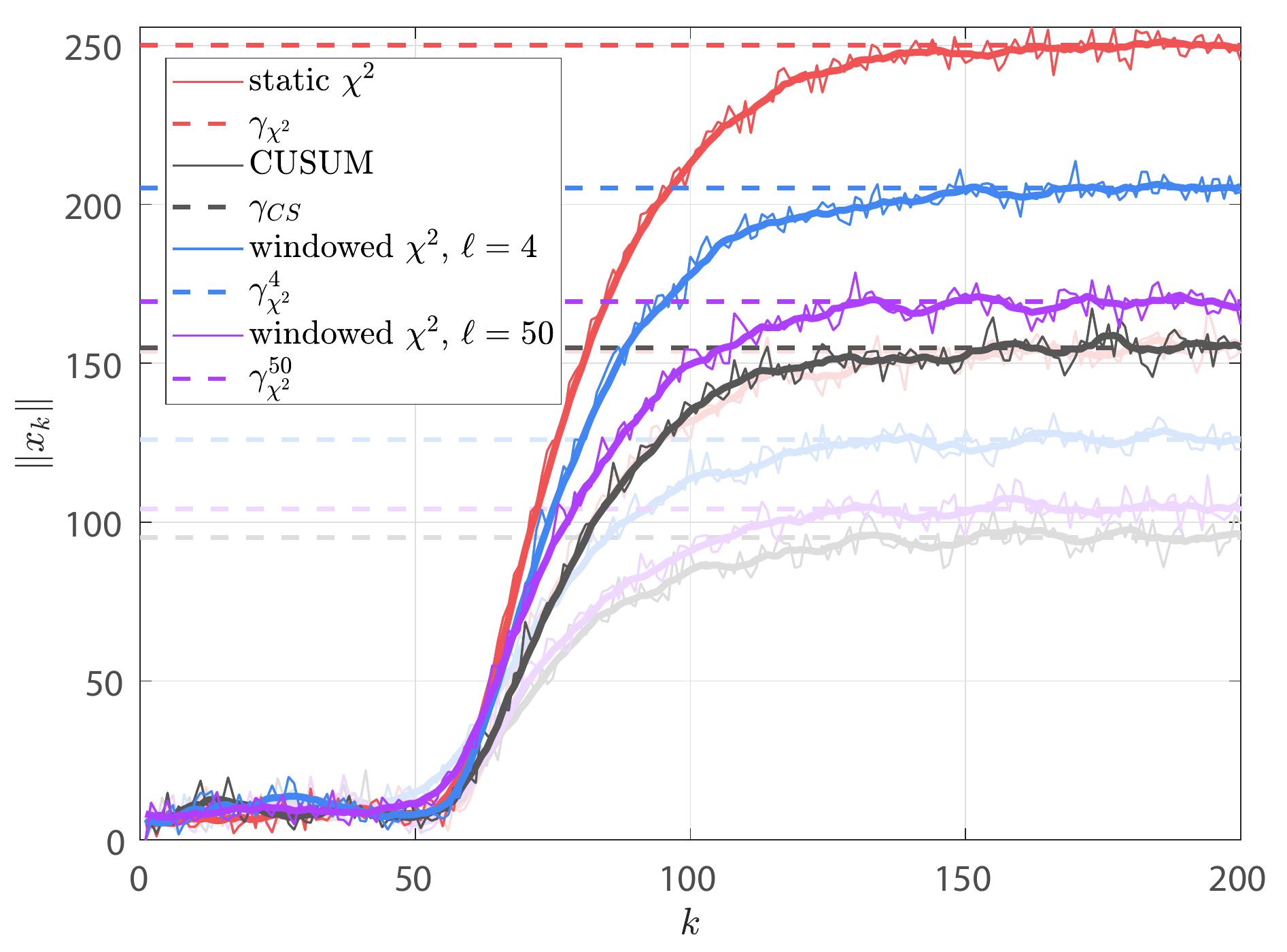}
\caption{Degradation of $||x_k||$ due to zero-alarm attacks. Attacks are induced at k = 51. Opaque curves denote the worst case zero-alarm attacks specified by Proposition \ref{prop:worst_attack}. Transparent curves denote zero-alarm attacks in which $\bar\delta=\mathbf{1}_{p\times 1}$. In both cases, the thicker line represents a moving average of the stochastic norm of the state using a moving window of 20 steps.}
\label{fig:state_deviation}
\end{figure}

\begin{figure*}[t]
\footnotesize
$F=\begin{pmatrix}
0.0273 & 0 & 0 & 0\\ 
0 & 0.0268 & 0.0001 & 0.0068 \\
0 & 0.0004 & 0 & 0.0018 \\
0 & 0.0619 & 0.0055 & 0.2478
\end{pmatrix},\ $
$G=\begin{pmatrix}
0.0271 & 0 & 0 \\ 
0 & 0.2665 & 0.0001 \\
0 & 0.0005 & 0.0276 \\
0 & 0.0761 & 0.0114
\end{pmatrix},\ $
$L=\begin{pmatrix}
0.0033 & 0 & 0 \\ 
0 & 0.0033 & 0 \\
0 & 0 & 0 \\
0 & 0.0147 & 101.3810
\end{pmatrix},
\ R_2 = 100 \times I_3,\ $

\vspace{2ex}

$C = \begin{pmatrix}
1 & 0 & 0 & 0 \\ 
0 & 1 & 0 & 0 \\
0 & 0 & 1 & 0
\end{pmatrix},$
$K=\begin{pmatrix}
3.2856 & -0.7139 & -0.8301 & 1.4940 \\ 
-0.0244 & 5.0912 & 2.0507 & -3.6645 \\
0.2707 & 54.5562 & 99.8275 & -117.5190
\end{pmatrix},\ $
$R_1 = \begin{pmatrix} 13.8785&0&0&0\\0&13.6531&0.0141&2.1122\\0&0.0141&1.3808&0.2623\\0&2.1122&2.623&34.1805 \end{pmatrix}. $

\vspace{1ex}
\hrule
%$\Sigma=\begin{pmatrix}
%113.8880 & 0 & 0\\ 
%0 & 113.6640 & 0.0147 \\
%0 & 0.0147 & 101.3810
%\end{pmatrix}$
\end{figure*}

It is worth noting that this result is system independent, and thus purely a comparison of the detectors in the limiting case. Considering the case when $p=1$, in Fig. \ref{fig:damage_vs_ell}, we show the nonlinear trade-offs between the selection of $\beta$ and $\ell$ to maintain the same false alarm rate $\mathcal{A}$. This plot demonstrates the convergence of the distribution as $\ell\to\infty$ to be centered at $p=1$ with diminishing variance.

The reader might notice that the chi-squared and windowed chi-squared detectors can achieve better-than-CUSUM values of state deviation by selecting a large false alarm rate ($\approx 40\%$ or larger). Using a detector with such high false alarm rates overlooks the practical issue of dealing with a large number of alarms. It also does not consider the effect an attacker can have if they execute attacks that leverage the part of the distribution beyond the threshold for their attacks. Zero-alarm attacks do not quantify the extra damage caused by such ``hidden'' attacks \cite{Carlos_Justin3}.

\subsection{Demonstration}
To validate our theoretical results, we use a model developed in \cite{Patton_Book,Wata} for a well stirred chemical reactor with heat exchanger. The state, input, and output vectors of the reactor are:
\begin{equation}
x(t) = \begin{pmatrix} C_o \\ T_o \\ T_w \\ T_m \end{pmatrix}, \hspace{1mm}
u(t) = \begin{pmatrix} C_u \\ T_u \\ T_{w,u} \end{pmatrix}, \hspace{1mm}
y(t) = \begin{pmatrix} C_o \\ T_o \\ T_w \end{pmatrix},
\end{equation}
$C_o$: Concentration of the chemical product.\\
$T_o$: Temperature of the product.\\
$T_w$: Temperature of the jacket of water of heat exchanger.\\
$T_m$: Coolant temperature.\\
$C_u$: Inlet concentration of reactant.\\
$T_u$: Inlet temperature.\\
$T_{w,u}$: Coolant water inlet temperature.\\[1mm]
The original nonlinear model is linearized about the origin $x(t) = 0_{4 \times 1}$ and the system matrices are given below. In Fig. \ref{fig:state_deviation}, we show the deviation of the state norm $\|x_k\|$ subject to attacks that begin at $k=51$. We show the result of two different sets of zero-alarm attack sequences. One set (in opaque) represents the deviation due to worst case attacks generated in accordance to Proposition \ref{prop:worst_attack}. The second set of attacks (semi-transparent) are zero-alarm attacks with $\bar\delta=\mathbf{1}_{p\times 1}$. All sensors are used for detection and the detectors are tuned for a false alarm rate of 5\%, $\mathcal{A}=0.05$. Following our tuning results, we select $\alpha=7.81$ (static chi-squared threshold), $\beta=21.03$ (windowed chi-squared threshold, $\ell=4$), $\beta=179.58$ (windowed chi-squared threshold, $\ell=50$), and $b=3$ (CUSUM bias; CUSUM threshold is not needed for these simulations, but for completeness it is $\tau=0.86$). These lead to predicted steady state errors shown in dashed lines in Fig. \ref{fig:state_deviation}.

In each case, the steady-state deviation represents the ``damage'' that an attacker can be confined to as long as the attacker wishes to raise no alarms (to hide from detection). As expected, the largest steady-state deviation is permitted by the static chi-squared detector. As the window length increases (to $\ell=4$ and $\ell=50$ in the figure), the deviation decreases approaching the smallest deviation, accomplished by the CUSUM detector. An attacker is most effective in dealing damage by selecting the worst case attacks described in Proposition \ref{prop:worst_attack}. Changing the definition of $\bar\delta$ has the same effect to the damage allowed by each detector (due to the similar structures of the respective $\gamma$ definitions); here the worst case attack causes 163\% higher damage than the naive attack with $\bar\delta=\mathbf{1}_{p\times 1}$.

%For higher window lengths ($\ell > 8$), the windowed chi-square leads to a smaller steady state deviation. As stated by the Propositions, given that $\rho{[F]}<1$, $lim_{k \to \infty} ||E[x_k]|| =\gamma_{CS} =135.15$ for CUSUM, and $lim_{k \to \infty} ||E[x_k]|| =\gamma_{x^2} =173.05$ for static chi-squared and $lim_{k \to \infty} ||E[x_k]|| =\gamma_{x^2}^{\ell} =17.20$.

\section{Conclusions}

In this paper, we have employed the windowed chi-square procedure for attack detection. For the case of zero alarm attacks, we identified the limiting steady-state state deviation allowed by each type of detector: windowed chi-squared, static chi-squared, and CUSUM. The window chi-squared provided insight into the role that the window length plays in dynamic detectors. In particular, we show that the performance of the windowed chi-squared detector approaches that of the CUSUM as the window length gets longer.

\addtolength{\textheight}{-12cm}   % This command serves to balance the column lengths

%%%%%%%%%%%%%%%%%%%%%%%%%%%%%%%%%%%%%%%%%%%%%%%%%%%%%%%%%%%%%%%%%%%%%%%%%%%%%%%%

%%%%%%%%%%%%%%%%%%%%%%%%%%%%%%%%%%%%%%%%%%%%%%%%%%%%%%%%%%%%%%%%%%%%%%%%%%%%%%%%



% \section*{ACKNOWLEDGMENT}

%%%%%%%%%%%%%%%%%%%%%%%%%%%%%%%%%%%%%%%%%%%%%%%%%%%%%%%%%%%%%%%%%%%%%%%%%%%%%%%%

\bibliographystyle{IEEEtran}
\bibliography{security}

\end{document}